\documentclass[preprint,number]{elsarticle}
\journal{Journal of Non-Newtonian Fluid Mechanics}
\biboptions{sort&compress}

\usepackage{graphicx} 
\usepackage{physics}
\usepackage[a4paper, total={6in, 8in}]{geometry}
\usepackage{xcolor}
\usepackage{multirow}
\usepackage{placeins}
\usepackage{subcaption}
\usepackage{amsmath,amsfonts,amssymb,stackengine}
\usepackage{mathtools}
\usepackage{marginnote}

\newcommand{\T}{\vb{T}}

\newcommand{\C}{\vb{C}}
\newcommand{\I}{\vb{I}}
\newcommand{\kx}{\hat k}
\newcommand{\ky}{\hat l}
\renewcommand{\u}{\vb{u}}

\begin{document}

\begin{frontmatter}



\title{The polymer diffusive instability in highly concentrated polymeric fluids}


\author[cambridge]{Theo Lewy}
\ead{tal43@cam.ac.uk}
\corref{cor1}
\cortext[cor1]{Corresponding Author}

\affiliation[cambridge]{organization={Department of Applied Mathematics and Theoretical Physics, University of Cambridge},
            city={Cambridge},
            postcode={CB3 0WA},
            country={United Kingdom}}

\author[cambridge]{Rich Kerswell}

\begin{abstract}
The extrusion of polymer melts is known to be susceptible to `melt fracture' instabilities, which can deform the extrudate, or cause it to break entirely. Motivated by this, we consider the impact that the recently discovered polymer diffusive instability (PDI) can have on polymer melts and other concentrated polymeric fluids using the Oldroyd-B model with the effects of polymer stress diffusion included. Analytic progress can be made in the concentrated limit (when the solvent-to-total-viscosity ratio $\beta \rightarrow 0$), illustrating the boundary layer structure of PDI, and allowing the prediction of its eigenvalues for both plane Couette and channel flow. We draw connections between PDI and the polymer melt `sharkskin' instability, both of which are short wavelength instabilities localised to the extrudate surface. Inertia is shown to have a destabilising effect, reducing the smallest Weissenberg number ($W$) where PDI exists in a concentrated fluid from $W\sim 8$ in inertialess flows, to $W \sim 2$ when inertia is significant.
\end{abstract}



\begin{keyword}
Elastic Instability \sep Viscoelastic Flow \sep Polymer Melt \sep Melt Fracture



\end{keyword}

\end{frontmatter}
\section{Introduction}

The extrusion of plastic is a widespread industrial process in which plastics are shaped for countless applications. Plastic is melted and then forced through a slit (the `die'), where it can then be shaped into sheets, tubes or moulded precisely. This molten plastic is a type of viscoelastic `polymer melt', in which the fluid consists of long chain polymers, causing the fluid to feel an added force due to polymer stress.

It is known that the extrusion of polymer melts is an unstable process, as the extrudate can become distorted, or break entirely in a process called `melt fracture' \citep{Denn,Denn3, Meulenbroek, Bertola,vinogradov}. This can place a limit on the speed of industrial plastic extrusion, slowing down the manufacturing process. Depending on the extrusion speed, different instabilities can be seen. The `sharkskin' instability is characterised by the extruded plastic developing a regular short-wavelength distortion, and it is thought that its origins lie with the interaction of the flow with the outlet of the channel \cite{origins_sharkskin, Graham}. The `spurt-flow' instability alternates between the sharkskin and smooth textures, and the proposed mechanism for this is due to wall-slip \cite{spurt_dynamics, black, Graham}. Even if these instabilities were suppressed, there is evidence that the pipe flow of polymer melts are intrinsically unstable due to a weakly non-linear subcritical bifurcation \cite{Meulenbroek, Bertola} that makes the system unstable to finite amplitude perturbations. This non-linear work was motivated by the apparent lack of a linear instability in polymer melts.

The recently discovered polymer diffusive instability (PDI) \cite{beneitez} is a linear instability that exists in an Oldroyd-B fluid and is localised at boundaries. It is elastic in origin, existing at vanishing Reynolds number, and requires the presence of polymer stress diffusion in the model. PDI provides a potential mechanism for the transition to elastic turbulence, a chaotic state characterised by a wide range of length scales in the absence of inertia, in planar geometries. The mechanism for elastic turbulence in curvilinear geometries is well understood - tension acts along curved streamlines which produces hoop stresses, generating a linear instability \cite{Shaqfeh}. However, in planar geometries there are no curved streamlines, and so this linear instability does not exist. Consequently, the process responsible for the transition to elastic turbulence in such geometries is currently under debate. Numerous mechanisms have been suggested, including linear instabilities \cite{beneitez, chaudhary_garg_subramanian_shankar_2021, garg, khalid_chaudhary_garg_shankar_subramanian_2021}, finite-amplitude perturbations \cite{Morozov, Morozov2} and transient growth effects \cite{jovanovic, jovanovic2, page_zaki_2014, page_zaki_2015, hariharan}. PDI represents one possible mechanism, and it has already been shown to trigger a transition from plane Couette flow to a sustained 3D chaotic state in a FENE-P fluid \cite{beneitez}.


While previous work has investigated PDI in dilute fluids \cite{beneitez}, we consider the concentrated limit of an Oldroyd-B fluid (where the solvent-to-total-viscosity ratio $\beta \rightarrow 0$). This corresponds to highly concentrated polymeric solutions, as well as polymer melts. 
%
%
In §2 we examine the governing equations, and demonstrate that analytic progress can be made when $\beta \rightarrow 0$ for both plane Couette flow and channel flow. 
We find the dispersion relation, identify the relevant scalings and detail the boundary layer structure present in PDI. In §3 we compare the analytic predictions of the growth rate to numerical results for plane Couette flow. We consider both inertialess and inertial flows as $\beta \rightarrow 0$, and find that the growth rate (in units of the applied shear) diverges. In this limit, the smallest Weissenberg number, W,  at which the system is unstable is $W_{crit} \sim 8$ for inertialess flows, and $W_{crit} \sim 2$ when inertia is significant. Numerically, we consider Weissenberg numbers of $1 \lesssim W \lesssim 1,000$, and Reynolds numbers $0 \leq Re \lesssim 100,000$ as well as solvent to total viscosity ratios $0.001 \lesssim \beta$. The order of magnitude of this last parameter is motivated by the use of the UCM fluid to model melt fracture \cite{Meulenbroek, black, Graham, Shore}, which assumes that $\beta$ is sufficiently small that it can be neglected. In section §4 we check the robustness of PDI by changing the stress boundary conditions from vanishing diffusion to the Neumann condition of vanishing stress flux. 
Finally, there is a discussion in §5 in which we draw connections to the phenomena of melt fracture.


\section{Formulation}

%
We consider the flow of an Oldroyd-B fluid in 2D, with flow direction $\vb{\hat x}$ and wall normal direction $\vb{\hat y}$  in a channel either driven by boundary motion (plane Couette flow) or an applied pressure gradient (channel flow). The effects of polymer stress diffusion are included, motivated by its presence in viscoelastic simulations to regularise hyperbolic constitutive relationships \cite{SURESHKUMAR199553, Sureshkumar1997, dubief1, Page1, QIN202326, OHTA2022104924}.

\subsection{Governing Equations}

The non-dimensional governing equations (e.g. \cite{beneitez}) are 
\begin{equation}
\begin{gathered}
    Re (\frac{\partial \u}{\partial t} + \u \cdot \nabla \u) = - \nabla p +  (1-\beta) \nabla \cdot \vb{T} + \beta \nabla^2 \vb{u},
\end{gathered}\label{governing:eq1}
\end{equation}
\begin{equation}
\begin{gathered}
    \frac{\partial{\C}}{\partial t} + \u \cdot \nabla \C - \nabla \u^T  \cdot \C - \C \cdot \nabla \u + \T = \varepsilon \nabla ^2 \C,
\end{gathered}\label{governing:eq2}
\end{equation}
\begin{equation}
\begin{gathered}
    \nabla \cdot \u = 0
\end{gathered}\label{governing:eq3}
\end{equation}
where $\T$ and $\C$ are the polymeric stress tensor and conformation tensor respectively. In the \mbox{Oldroyd-B} model they are simply related as follows
\begin{equation}
\begin{gathered}\label{governing:eq4}
\T = \frac{1}{W}(\C-\I).
\end{gathered}
\end{equation}
In the above, we have scaled each quantity using the channel half-height $h$, a characteristic speed $U_0$, and the total viscosity $\mu = \mu_p + \mu_s$ which sums the polymer and solvent viscosities ($\mu_p$ and $\mu_s$ respectively). In plane Couette flow, $U_0$ is taken to be the speed of the plates ($\u(x,y=\pm h,t) = \pm U_0\vb{\hat x} $) while in channel flow $U_0$ is defined in terms of the imposed pressure gradient along the channel ($U_0 = -h^2 / \mu \, dP/dx$). There are 4 dimensionless parameters after scaling,

$$ Re \coloneqq \frac{\rho   U_0 h}{\mu}, \qquad W \coloneqq \frac{\lambda   U_0}{h},$$
$$\beta \coloneqq \frac{\mu_s}{\mu}, \qquad \varepsilon \coloneqq \frac{\delta}{U_0 h},$$
where additionally we define $\rho$ as the fluid density, $\lambda$ as the relaxation timescale of the polymer, and $\delta$ as the polymer stress diffusivity. These dimensionless parameters are the Reynolds number $Re$, the Weissenberg number $W$, the solvent-to-total-viscosity ratio $\beta$ and the polymer stress diffusion coefficient $\varepsilon$.

\subsection{Boundary Conditions}

All equations will be solved subject to no-slip and no-penetration conditions on the velocity at the boundaries, yielding four boundary conditions for our 2D system. 
$$\begin{cases}
\u |_{y=\pm 1} =  \pm \vb{\hat x}   & \text{ Plane Couette flow}\\
\u |_{y=\pm 1} = \vb 0. & \text{ Channel flow}
\end{cases} $$
While there is evidence that wall-slip occurs for polymer melts \cite{Denn_wall_slip, spurt_dynamics, black, Graham, Denn}, it is not required for the existence of PDI, and so we do not include this in our model. We consider two types of boundary conditions on the polymer stress:
$$\begin{cases}
\nabla^2 \T |_{y=\pm 1} = \vb0 & \text{ Type I}\\
\frac{\partial \T}{\partial y}|_{y=\pm 1} = \vb0. & \text{ Type II}
\end{cases} $$
The Type I condition corresponds to the polymer stress not diffusing at the walls. This is equivalent to the more typical, but less concise condition of applying equation (\ref{governing:eq2}) with $\varepsilon=0$ at the wall \cite{SURESHKUMAR199553, Sureshkumar1997}, provided $\T$ has a continuous second  derivative. The Type II Neumann condition instead corresponds to the stress flux vanishing at the walls. While the Type I condition is physically more relevant, we introduce the Type II condition as a way to check robustness to the choice of boundary conditions. We focus on Type I conditions, and show how the results change in §\ref{robustness} when type II conditions are used instead.

\subsection{Base States}\label{base_states}

To find the base states we solve equations (\ref{governing:eq1})-(\ref{governing:eq4}) with $\u = U(y)\vb{\hat x}$, $\T = \T(y)$ and $p = P(x)$, subject to the above boundary conditions.

\subsubsection{Plane Couette flow}
We impose $dP/dx = 0$ for plane Couette flow. The solutions for both Type I and Type II stress boundary conditions are
$$ T_{xx} = 2 W, \qquad T_{xy} = 1, \qquad T_{yy} = 0, \qquad U=y.$$
This is independent of the stress boundary conditions as this solution has a constant stress.
\subsubsection{Channel flow}
A pressure gradient of $dP/dx = -1$ is imposed for channel flow. The laminar base state is altered by the presence of polymer diffusion. For Type I boundary conditions the exact solutions are
$$ T_{xx} = 2Wy^2 + 4\varepsilon W^2 \left(1 - \frac{\cosh(y/\sqrt{\varepsilon W})}{\cosh(1/\sqrt{\varepsilon W })}\right), \qquad T_{xy} = -y, \qquad T_{yy} = 0, \qquad U=\frac{1}{2}(1-y^2).$$
We can see this base flow contains a boundary layer with lengthscale $\sqrt{\varepsilon W}$. For Type II conditions we find a narrower boundary layer of lengthscale $\sqrt{ \varepsilon \beta W}$, with 
\begin{align*}
 T_{xx} &= 2Wy^2 + 4 \varepsilon W^2 \\ 
 & - \frac{2\varepsilon \beta W^2}{\cosh(1/\sqrt{\varepsilon \beta W})}\frac{(1-2\beta)}{(1-\beta)^2}\left((1-\beta)\frac{y}{\sqrt{\varepsilon \beta W}}\sinh(y/\sqrt{\varepsilon \beta W}) - 2\cosh(y/\sqrt{\varepsilon \beta W}) \right) \\
 & + \frac{\varepsilon W^2 (1-\beta) }{\cosh^2(1/\sqrt{\varepsilon \beta W})} \left(\frac{\beta}{4-\beta} \cosh(2y/\sqrt{\varepsilon \beta W }) + 1\right) \\
 & - \frac{2W}{(1-\beta)^2} \frac{\sqrt{\varepsilon W }}{\sinh(1/\sqrt{\varepsilon W })}\left((1-\beta) - 
  \sqrt{\varepsilon \beta W}\frac{(\beta^2+11\beta-6)}{(4-\beta)} \tanh(1/\sqrt{\varepsilon \beta W})               \right)
 \cosh(y/\sqrt{W \varepsilon}),\\
 T_{xy} &= -y + \sqrt{\varepsilon \beta W }\frac{\sinh(y/\sqrt{\varepsilon \beta W })}{\cosh(1/\sqrt{\varepsilon \beta W})}, \\
 T_{yy} &= 0, \\
  U &= \frac{1}{2}(1-y^2) - (1-\beta) \varepsilon W \left(\frac{\cosh(y/\sqrt{\varepsilon \beta W})}{\cosh(1/\sqrt{\varepsilon \beta W})}-1\right).
\end{align*}

\begin{figure}[ht!] 
\centering
    \begin{subfigure}[b]{0.33\textwidth} 
\includegraphics[width=\textwidth]{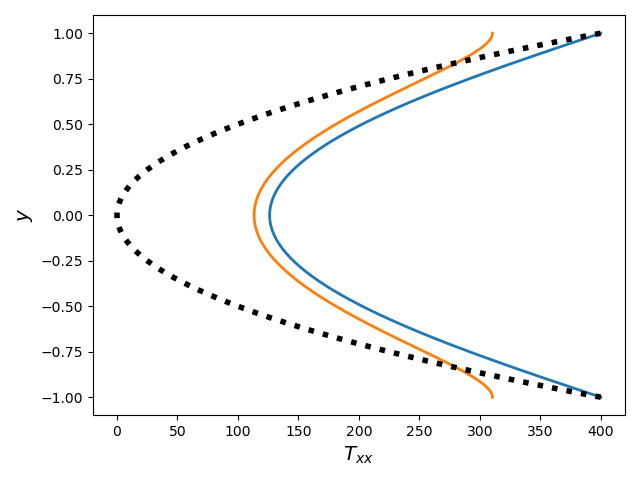}
\caption{}
  \end{subfigure}
    \begin{subfigure}[b]{0.32\textwidth} 
\includegraphics[width=\textwidth]{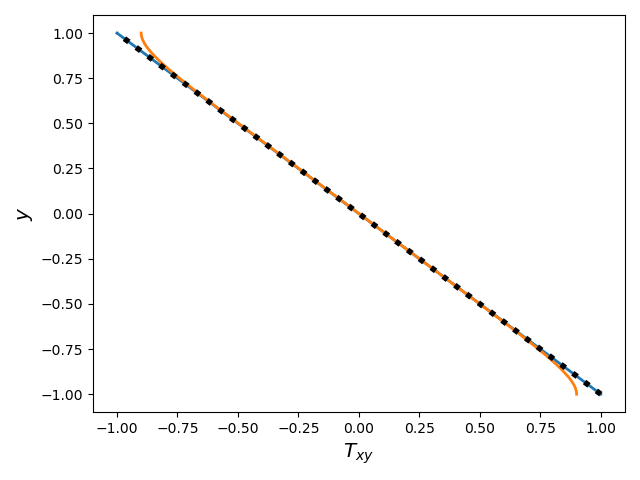}
\caption{}
  \end{subfigure}
  \begin{subfigure}[b]{0.33\textwidth} 
\includegraphics[width=\textwidth]{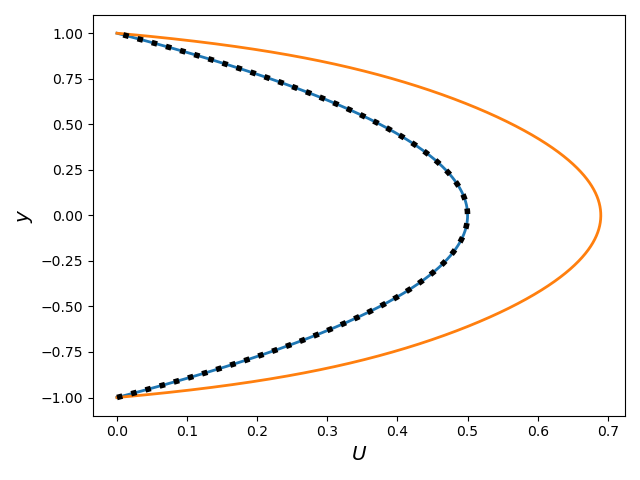}
\caption{}
  \end{subfigure}
\caption{The channel flow base state showing a) $T_{xx}$ b) $T_{xy}$ and c) $U$ with boundary conditions of no diffusion (blue) and no stress-flux (orange), plotted alongside the base flow obtained when $\varepsilon=0$ (black dotted). Parameters are $W=200, \varepsilon=0.001, \beta=0.05$ and $Re=0$. This shows how the presence of polymer stress diffusion can modify the base flow in channel flow.} \label{channel-base-flow}
\end{figure}

All of these laminar base flows were checked by comparing the analytic results to the numerical results obtained using the open-source python package Dedalus and its non-linear boundary value problem solver \cite{dedalus}. Note that the presence of boundary layers due to polymer stress diffusion could provide a way to verify the validity of this model experimentally. Both boundary conditions for channel flow have boundary layers of lengthscale $\sqrt{ \varepsilon W}$, which can be sizable. For example, a long chain polymer melt with $\varepsilon = 0.001$ and $W=100$ has a boundary layer of length $\sim 0.3 h$ or 15\% of the channel width. See figure \ref{channel-base-flow} for how the flow variables can be modified due to polymer stress diffusion.

%
%

\subsection{Linearised Equations}

We now linearise equations (\ref{governing:eq1})-(\ref{governing:eq4}) about the base states $u=U(y)\hat{\textbf{x}}$ and $\T = \vb T(y)$ from \S\ref{base_states}. We introduce the polymeric stress perturbation $\boldsymbol \tau$ and the streamfunction perturbation $\psi$, and assume $\boldsymbol \tau, \psi \propto e^{ik(x-ct)}$ where $c \in \mathbb{C}$ is an eigenvalue to be determined. This results in
$$\T = \vb T(y) + \boldsymbol \tau(y)e^{ik(x-ct)}, \qquad \vb u = U(y)\hat{\vb x} +     \begin{pmatrix}
         - D \psi(y)\\
        ik \psi(y)
    \end{pmatrix}e^{ik(x-ct)}$$
where $ D := \frac{d}{d y}$. Choosing to use the streamfunction formulation immediately satisfies the incompressibility equation (\ref{governing:eq3}). We take the curl of (\ref{governing:eq1}) to eliminate the pressure, and then linearise it to give
\begin{equation}\label{governing:eq6}
\begin{gathered}
 ik Re\left[(U-c)( D^2 - k^2) \psi - { D^2 U \psi} \right]  = 
- (1-\beta)\left[ ik D \tau_{xx} - ik D \tau_{yy} +  (D^2 + k^2) \tau_{xy}\right] + \beta( D^2 - k^2)^2\psi
\end{gathered}
\end{equation}
which is the viscoelastic analogue to the Orr-Sommerfeld equation \cite{orszag_1971}.
Using (\ref{governing:eq2}) and (\ref{governing:eq4}), we then obtain the linearised constitutive equation for the stress perturbation as
\begin{equation}\label{governing:eq5}
\begin{gathered}
    ik(U-c)
    \begin{pmatrix}
         \tau_{xx} &  \tau_{xy} \\
         \cdot &  \tau_{yy}
    \end{pmatrix}
        +
        ik\psi
    \begin{pmatrix}
         D T_{xx}&  D T_{xy} \\
         \cdot&  D T_{yy}
    \end{pmatrix}
    - DU
        \begin{pmatrix}
        2 \tau_{xy}&  \tau_{yy} \\
         \cdot& 0
    \end{pmatrix}   \\
    +
    \begin{pmatrix}
        2T_{xx}ik D\psi + 2T_{xy}  D^2 \psi & T_{xx}k^2\psi  + T_{yy} D^2 \psi \\
        \cdot & 2T_{xy}k^2 \psi - 2T_{yy}ik D \psi \\
    \end{pmatrix}
    +
    \frac{1}{W}
            \begin{pmatrix}
        2ik D\psi & k^2\psi +  D^2 \psi \\
        \cdot & -2ik D \psi \\
    \end{pmatrix}
    \\
    +
    \frac{1}{W}
     \begin{pmatrix}
         \tau_{xx} &  \tau_{xy} \\
         \cdot&  \tau_{yy}
    \end{pmatrix}
    = 
    \varepsilon( D^2 - k^2)      \begin{pmatrix}
         \tau_{xx} &  \tau_{xy} \\
         \cdot&  \tau_{yy}
    \end{pmatrix}
\end{gathered}
\end{equation}
where we suppress the lower off diagonal element for clarity as all tensors are symmetric. The analysis for plane Couette flow and channel flow with vanishing stress diffusion at the boundary is very similar and so we keep all equations sufficiently general to cover both systems.

\subsection{Asymptotics}

We now rescale our variables so that we can identify which terms in (\ref{governing:eq6}) and (\ref{governing:eq5}) can be neglected in the polymer melt limit. To motivate our scaling, we observe that for inertialess flow $c \sim \sqrt{{\varepsilon W (1-\beta)}/{\beta}}$ and $k \sim \sqrt{{W (1-\beta)}/{\varepsilon\beta}}$. Then, noting PDI is localised at the boundaries \cite{beneitez}, we introduce a boundary variable $\hat y$ at the lower boundary, and rescale quantities as follows
$$
\hat k = \sqrt{\frac{\varepsilon\beta}{W (1-\beta)}} k, \qquad
\hat{y}=  \sqrt{\frac{W (1-\beta)}{\varepsilon\beta}}(y+1), \qquad
\hat{c} = \sqrt{\frac{\beta}{\varepsilon W (1-\beta)}}\left(c-U(-1)\right),
$$
$$ \hat{ \tau}_{xx}= \tau_{xx}, \qquad 
\hat{ \tau}_{xy}= \tau_{xy}, \qquad 
\hat{ \tau}_{yy}=W \tau_{yy}, \qquad 
\hat{\psi} = \frac{W}{\varepsilon}\psi. $$
Using these rescalings, as well as the Taylor expansion of $U$ around $y=-1$
\begin{equation}
    U-c =  - \sqrt{\frac{\varepsilon W (1-\beta)}{\beta}} \hat c + \sqrt{\frac{\varepsilon\beta}{W (1-\beta)}} \hat y + \ldots
\end{equation}
equations (\ref{governing:eq6}) and (\ref{governing:eq5}) become
\begin{align}\label{rescaled:eq2}
 \frac{\varepsilon Re}{\beta} i \hat{k} 
 \left[
 \left(\frac{\beta}{W(1-\beta)} \hat{y} - \hat{c} \right)(\hat{D}^2 -\hat{k}^2) \hat{\psi} - \sqrt{\varepsilon}\left(\frac{\beta}{W(1-\beta)}\right)^{3/2} { D^2 U \hat{\psi}} 
 \right] \nonumber\\
 = 
-\left(i\hat k  \hat D \hat{\tau}_{xx} - \frac{1}{W}i\hat{k} \hat D \hat{\tau}_{yy} +  (\hat{D}^2 + \hat{k}^2) \hat{\tau}_{xy}\right) + ( \hat{D}^2 - \hat{k}^2)^2 \hat{\psi}
\end{align}
and
\begin{equation}\label{rescaled:eq1}
\begin{gathered}
i\hat k \left(\frac{\beta}{W(1-\beta)} \hat y - \hat c \right)
    \begin{pmatrix}
         \hat \tau_{xx} &  \hat  \tau_{xy} \\
         \cdot &  \hat  \tau_{yy}
    \end{pmatrix}
        +
        \sqrt{\frac{\varepsilon \beta}{W(1-\beta)}} i\hat k \hat \psi
    \begin{pmatrix}
          \frac{1}{W} DT_{xx}&  \frac{1}{W}DT_{xy} \\
         \cdot& \frac{1}{W} DT_{yy}
    \end{pmatrix} \\
    - \frac{\beta}{W(1-\beta)} DU
        \begin{pmatrix}
        2 \hat \tau_{xy}&  \frac{1}{W} \hat \tau_{yy} \\
         \cdot& 0
    \end{pmatrix}   
    + \frac{1}{W}
     \begin{pmatrix}
        2T_{xx}i\hat k \hat D \hat \psi + 2T_{xy} \hat  D^2 \hat \psi & T_{xx} \hat k^2 \hat \psi  + T_{yy} \hat D^2 \hat \psi \\
        \cdot & 2W T_{xy}\hat k^2 \hat \psi - 2WT_{yy}i\hat k\hat  D \hat \psi \\
    \end{pmatrix} \\
    +
    \frac{1}{W^2}
            \begin{pmatrix}
        2i\hat k \hat  D \hat  \psi & \hat  k^2 \hat  \psi +  \hat  D^2 \hat  \psi \\
        \cdot & -2W i\hat  k \hat  D \hat  \psi \\
    \end{pmatrix}
    +
     \frac{\beta}{W(1-\beta)}
     \frac{1}{W}
     \begin{pmatrix}
         \hat \tau_{xx} &  \hat \tau_{xy} \\
         \cdot&  \hat \tau_{yy}
    \end{pmatrix} = (\hat D^2 - \hat k^2)      \begin{pmatrix}
         \hat \tau_{xx} & \hat \tau_{xy} \\
         \cdot& \hat \tau_{yy}
    \end{pmatrix}.
\end{gathered}
\end{equation}
In the limit of $\beta/W(1-\beta) \rightarrow 0$ for finite $\varepsilon Re/\beta$ and $W$ these become
\begin{equation}\label{rescaled:eq4}
\begin{gathered}
\frac{\varepsilon Re}{\beta} i \kx \hat {c}(  {\hat D}^2 -  \kx^2 ) {\hat\psi} + (  {\hat D}^2 -  \kx^2)^2 {\hat\psi} - i \kx  {\hat D} (\hat{ \tau}_{xx} - \frac{\hat \tau_{yy}}{W}) -   ({\hat  D}^2 + \kx^2) \hat { \tau}_{xy} = 0
\end{gathered}
\end{equation}
\begin{equation}\label{rescaled:eq3}
\begin{gathered}
    (-i \kx \hat c +  \kx^2 -   {\hat  D}^2)
    \begin{pmatrix}
         \hat{ \tau}_{xx}&  \hat{ \tau}_{xy} \\
         \cdot&  \hat{ \tau}_{yy}
    \end{pmatrix}
        +  \begin{pmatrix}
        4i \kx  {\hat D}{\hat\psi} + \frac{2}{W}  {\hat D}^2  {\hat\psi} + \frac{2}{W^2}i \kx  {\hat D}{\hat\psi} & 2  \kx^2{\hat\psi}  + \frac{1}{W^2} \kx^2{\hat\psi} + \frac{1}{W^2}  {\hat D}^2  {\hat\psi}\\
        \cdot & 2 \kx^2  {\hat\psi} - \frac{2}{W} i\kx  {\hat D}  {\hat\psi} \\
    \end{pmatrix} 
    = 
    0.
\end{gathered}
\end{equation}
This derivation makes use of the fact that for both plane Couette and channel flow,
$$\vb T(y) = \vb T(-1) + O(y+1)   = \vb T(-1) + O(\sqrt{\frac{\varepsilon\beta}{W (1-\beta)}}\hat y) \rightarrow \vb T|_{-1} =      \begin{pmatrix}
         2W&   1 \\
          \cdot &  0
    \end{pmatrix}$$
allowing us to treat the base stress as constant in the region where PDI is active. The same simplification also occurred for the base velocity $U$, allowing $U-c$ to be treated as constant in the given limit. This is now a system with constant coefficients. Applying the operator $ (-i \kx \hat{c} +  \kx^2 -   \hat{ D}^2)$ to equation (\ref{rescaled:eq4}), and then using equation (\ref{rescaled:eq3}) to rewrite each $\hat  \tau$ in terms of $ \hat \psi$ gives
\begin{equation}
    \left[ (\hat D^2 - \kx ^2)^2 + i \kx   \hat c \left(1+\frac{\varepsilon Re}{\beta}\right)(\hat D^2 - \kx^2) + \left(\kx - \frac{i}{W}\hat D \right)^2 + \left(1+\frac{1}{W^2} - \frac{\varepsilon Re}{\beta}   \hat c^2\right)  \kx^2\right]  (  \hat D^2 - \kx ^2) \hat \psi = 0.
\end{equation}
This admits solutions of the form $\hat \psi =   \hat \psi_0 e^{i \ky   \hat y}$, where $\ky$ satisfies the dispersion relation
\begin{equation} \label{dispersion} 
    \left[(\kx^2 + \ky ^2)^2 - i \kx   \hat c \left(1+\frac{\varepsilon Re}{\beta}\right)(\kx^2 + \ky^2) + \left(\kx + \frac{\ky}{W}  \right)^2 + \left(1+\frac{1}{W^2} - \frac{\varepsilon Re}{\beta}   \hat c^2\right)  \kx^2\right]  (\kx^2 + \ky ^2) \hat \psi_0 = 0
\end{equation}
which relates the complex eigenvalue $\hat c$ to the spatial wavenumber pair ($\hat k$, $\hat l$). We now consider the superposition of these solutions in order to satisfy the boundary conditions.

\subsection{Boundary Conditions} 
For a given $\hat c$ and $\hat k$, equation (\ref{dispersion}) yields 6 possible $\hat l$ that can contribute to the eigenmode, yielding 
\begin{equation}
\hat \psi = \sum^6_{n=1} \hat \psi_{0,n} e^{i\hat l_n \hat y}.
\end{equation} 
We are considering PDI localised to the lower boundary and so we can discard all $\hat l$ that do not decay as $\hat y \rightarrow \infty$. This leaves only $\ky$ with positive imaginary parts. One of these is an exact solution, $\ky = i\kx$, which appears as PDI is acting on the vorticity field, $(\hat D^2 - \kx^2)\psi = \hat \nabla^2 \psi$, and this component of the eigenmode has zero vorticity. We assume two more $\ky$ exist with positive imaginary part, $\ky_+,  \ky_-$.
We therefore have
\begin{equation}
      \hat \psi = Ae^{- \kx   \hat y} + B_+e^{ i \ky_+  \hat y} + B_-e^{ i \ky_- \hat  y}.
\end{equation}
Using the no-penetration and no-slip conditions results in $B_\pm = A f_\pm (\hat k, \hat l_+, \hat l_-)$ where $f_\pm$ can be written analytically. We can then substitute the resultant $ \hat \psi$ into equation (\ref{rescaled:eq3}) to obtain a 2nd order differential equations for each component of $  \hat  \tau$. 
\begin{equation}
       (-i \kx \hat c +  \kx^2 -   {\hat  D}^2)\boldsymbol {\hat{\tau}} = A\vb g_1(\hat k, W) e^{- \kx  \hat  y} + A\vb g_2(\hat k, \hat l_+, \hat l_-, W) e^{ i \ky_+ \hat   y} + A\vb g_3(\hat k, \hat l_+, \hat l_-, W)  e^{ i \ky_-  \hat  y}
\end{equation}
where the $\vb g_i$ can be found exactly. Solving this differential equation introduces a new scale in the stress field, with decay rate $\pm \sqrt{ \kx^2 - i  \kx   \hat c}$ (the sign of this is set so that the resultant term decays in the far field). To simplify our notation, we will define the complex square root to pick the solution with a positive real part. Using the Type I stress boundary conditions then sets the coefficient of this introduced complementary function so that 
\begin{align}
       \boldsymbol {\hat{\tau}} = A\vb h_1(\hat k, \hat c, W) e^{- \kx  \hat  y} + A\vb h_2(\hat k, \hat l_+, \hat l_-, \hat c, W) e^{ i \ky_+ \hat   y} + A\vb h_3(\hat k, \hat l_+, \hat l_-, \hat c, W)  e^{ i \ky_-  \hat  y} \nonumber \\ + A\vb h_4(\hat k, \hat l_+, \hat l_-, \hat c, W) e^{-\sqrt{ \kx^2 - i  \kx \hat  c} \, \hat  y}
\end{align}
where the $\vb h_i$ can be found analytically. We substitute $\hat \psi$ and $\boldsymbol {\hat{\tau}}$ into equation (\ref{rescaled:eq4}), and consider the resultant coefficients of each exponential on the left-hand side. The coefficient of $\exp(-\hat k \hat y)$ trivially vanishes, while those of $\exp(i \hat l_\pm \hat y)$ just recover that the dispersion relation (\ref{dispersion}) is satisfied for $\ky_\pm$. Examining the coefficient of $\exp(-\sqrt{ \kx^2 - i  \kx \hat  c} \, \hat  y)$ however gives a further condition. For the cases of plane Couette and channel flow with Type I stress boundary conditions, this extra condition is
\begin{align}\label{discretisation}
\mathcal{L} :=  
\frac{(i  \ky_+ -  \kx)}{( i \kx   \hat c - \kx^2  -  \ky_+ ^2)} \left[2 \kx \left(2(1+\frac{1}{W^2}) \kx  \ky_+ + \frac{\kx^2 +  \ky_+^2}{W}\right) 
 - i \frac{2 \kx^2 - i  \kx  \hat  c}{\sqrt{ \kx^2 - i  \kx  \hat  c}} (2 \kx^2 + \frac{ \kx ^2 -  \ky_+^2}{W^2})   \right] \nonumber \\
 - 
\frac{(i  \ky_- -  \kx)}{( i \kx   \hat c - \kx^2  -  \ky_- ^2)} \left[2 \kx \left(2(1+\frac{1}{W^2}) \kx  \ky_- + \frac{\kx^2 +  \ky_-^2}{W}\right) 
 - i \frac{2 \kx^2 - i  \kx  \hat  c}{\sqrt{ \kx^2 - i  \kx  \hat  c}} (2 \kx^2 + \frac{ \kx ^2 -  \ky_-^2}{W^2})   \right] = 0.
\end{align}
Solving equations (\ref{dispersion}) and (\ref{discretisation}) together gives the eigenvalue $\hat c_{pred}$. This was done by using the Newton-Raphson method to find zeros of $\mathcal{L}(\hat c)=\mathcal{L}(\hat c, \ky_+(\hat c), \ky_-(\hat c))$, where $\ky_\pm(\hat c)$ are the solutions to equation (\ref{dispersion}) with positive imaginary part.

In summary, the asymptotic limit pursued here has turned the original system with 4 parameters $Re, \beta, \varepsilon$ and $W$ into a constant-coefficient eigenvalue problem with only 2 parameters $W$ and $\varepsilon Re/\beta$, i.e.
\begin{equation}
\hat c_{pred} \,\, = \, \,
\hat c_{pred}\left(\hat k; \,W,\,\frac{\varepsilon Re}{\beta} \right) 
\,\, = 
\lim_{\frac{\beta}{W(1-\beta)}\,\to \,0} 
\hat c \left(\hat k; \,W,\,\frac{\varepsilon Re}{\beta}, \,\frac{\beta}{W(1-\beta)}, \,\varepsilon \right).
\label{reduction}
\end{equation}
In particular, the parameter $\varepsilon$ now only appears in the parameter group $\varepsilon Re / \beta$ which measures the strength of inertia.






\section{Results}\label{results}

We now compare our analytic predictions to numerical computations in the polymer melt limit. All numerical results are computed using the open-source python package Dedalus \cite{dedalus}. As the results for plane Couette and channel flow are similar, we discuss plane Couette in the main text and place the key results for channel flow in \ref{analogous_results}.  We firstly consider creeping flows ($Re = 0$), and then investigate how this flow is altered due to the presence of inertia ($Re > 0$).

\subsection{Inertialess flows: $Re = 0$}
\begin{figure}[h!] 
\centering
  \begin{subfigure}[b]{0.49\textwidth}
    \centering
\includegraphics[width=\textwidth]{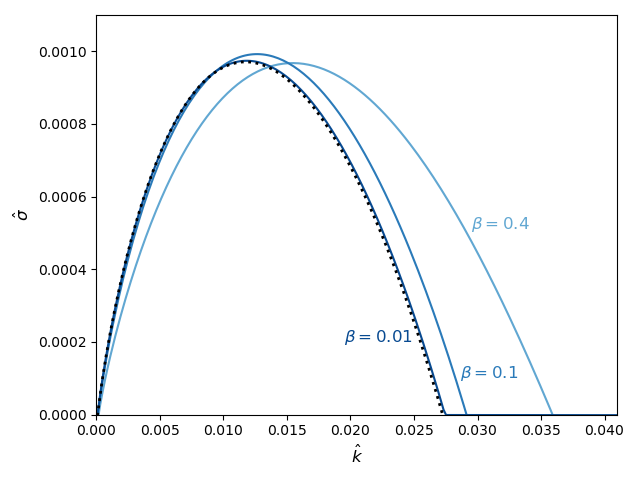}
    \caption{}
    \label{dispersion_beta}
  \end{subfigure}
  \begin{subfigure}[b]{0.49\textwidth}
    \centering
\includegraphics[width=\textwidth]{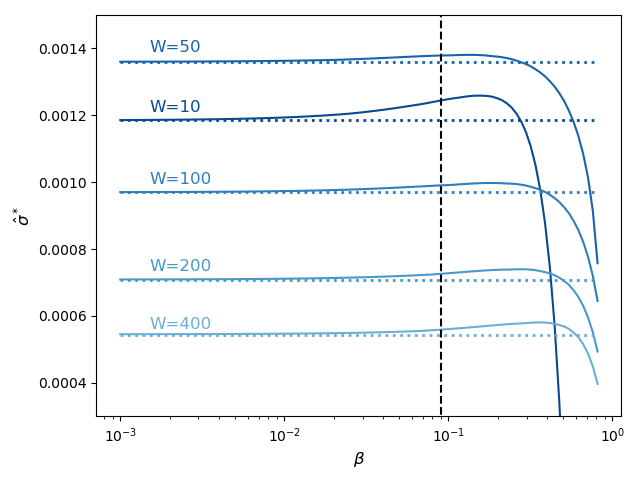}
    \caption{}
    \label{growth_rate_creeping}
  \end{subfigure}
\caption{a) Dispersion relations obtained from our analytic predictions (dotted lines) vs numerics (solid lines) for $W=100$, $Re=0$, $\varepsilon=0.0001$, $\beta = 0.01,0.1,0.4$. b) The scaled maximum growth rate for our analytic predictions (dotted lines) vs numerics (solid lines) at $\varepsilon=0.0001$, $Re=0$, $W =10,50,100,200,400$. The limits for these values of $W$ are $\hat \sigma ^* = 0.001185, 0.001360, 0.000971, 0.000709, 0.000545$ respectively. The dashed black line corresponds to $\beta_{lim}(Re=0) = 0.090$ which is discussed later in section \ref{validity}, and represents where the predictions are within $5\%$ of the numeric values for all $\beta < \beta_{lim}$. These plots show that the analytic theory agrees with the numerics in the polymer melt limit, and both $\hat \sigma $ and $\hat k$ are shown to be $O(1)$ as $\beta \rightarrow 0$.} 
\label{predictions_scaling}
\end{figure}

We begin by verifying the validity of the predicted scalings. To help with this we define the scaled and unscaled growth rates 
\begin{equation}
    \sigma \coloneqq \mathcal{R}(-i k c),  \qquad \hat \sigma \coloneqq \mathcal{R}(-i \hat k \hat c), 
\end{equation} 
where $\mathcal{R}$ denotes taking the real part. Hence
\begin{equation}
\sigma = \frac{W(1-\beta)}{\beta} \hat \sigma.
\end{equation} 
In addition, we define $\hat \sigma^*$ and $\sigma ^*$ to be the scaled and unscaled most unstable growth rates across all wavenumbers. Figure \ref{dispersion_beta} shows the dispersion relations of $\hat \sigma$ against $\hat k$ as $\beta$ varies, demonstrating that our analytic predictions match the numerics in the polymer melt limit. It also suggests that $\hat k = O(1)$ and $\hat \sigma = O(1)$ as $\beta \rightarrow 0$, verifying our predicted scalings. Figure \ref{growth_rate_creeping} shows that $\hat \sigma ^*$ tends to a constant, and that this limiting value of $\hat \sigma^*$ is non-monotonic in $W$. The maximum growth rate $\sigma^* = {W(1-\beta)}/{\beta} \,\hat \sigma^*$ therefore diverges at small $\beta$, as $\hat \sigma = O(1)$ as $\beta \rightarrow 0$. We emphasise that the timescale on which this divergence occurs is the inverse of the applied shear rate $h/U_0$ so this divergence is not a feature of the choice of timescale. Physically, this means that the instability becomes stronger the more concentrated the polymer melt is and suggests that PDI is particularly of interest for concentrated polymeric fluids.

\begin{figure}[h!] 
\centering

  \begin{subfigure}[b]{0.45\textwidth}
    \centering
\includegraphics[width=\textwidth]{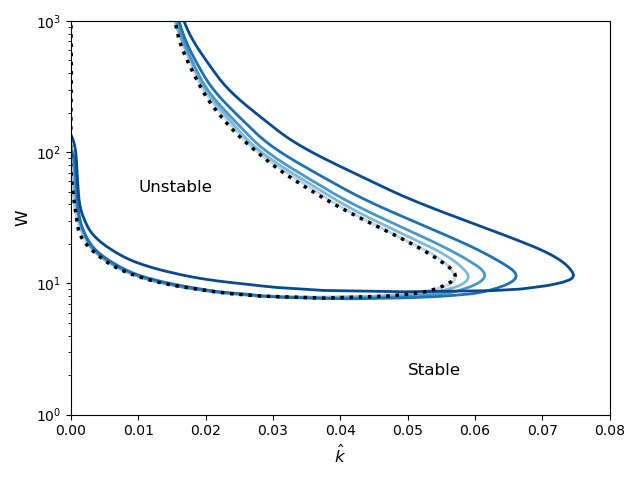}
    \caption{}
    \label{neutral_curve_beta}
  \end{subfigure}
    \begin{subfigure}[b]{0.45\textwidth}
    \centering
\includegraphics[width=\textwidth]{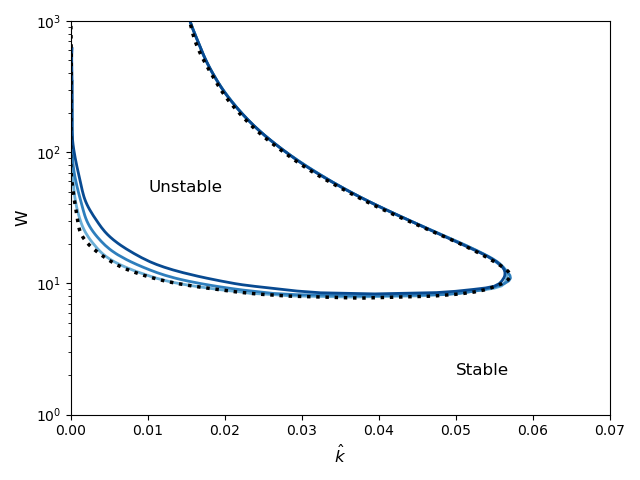}
    \caption{}
    \label{neutral_curve_eps}
  \end{subfigure}
\caption{Neutral stability curves in the ($W$, $\hat k$) plane with a) $\varepsilon=0.0001$ and, from lightest to darkest, $\beta = 0.01, 0.05, 0.1, 0.2, 0.4$  and b) $\beta =0.01$ and, from lightest to darkest, $\varepsilon = 0.1, 0.5, 1$. Dotted lines correspond to our analytic predictions, while solid lines show numeric neutral curves. These plots confirm the analytic predictions for the neutral curve.} \label{neutral_curves_creeping}
\end{figure}

Figure \ref{neutral_curves_creeping} shows the neutral stability curves in ($W$, $\hat k$) space when $\beta$ or $\varepsilon$ change. When $Re=0$, the derived analytic results show that $\hat c$ depends only on $W$ and $\hat k$ in the small $\beta/W(1-\beta)$ limit (that is, (\ref{reduction}) with $\varepsilon Re/\beta=0$), and hence the stability of the system is entirely shown by this plot. We show in figure \ref{neutral_curve_beta} how the neutral curve changes as $\beta$ is taken out of this limit. The analytic theory reproduces the $\beta = 0.01$ neutral curve almost exactly, and remains qualitatively accurate for viscosity ratios as large as $\beta=0.4$. Similarly, in figure \ref{neutral_curve_eps} we increase $\varepsilon$. The analytic theory is accurate for physically realistic values of polymer diffusion, with it remaining accurate for values as large as $\varepsilon=0.1$. The reason the neutral curve stops being accurate for larger $\varepsilon$ is due to the delocalisation of the eigenfunction near the boundary, when the analytic theory assumed localisation. The boundary layer thickness scales like $\sqrt{\varepsilon \beta / W(1-\beta)}$ and so when $\varepsilon$ is too large the eigenfunction interacts with the other wall.

The extrusion of plastics is typically modelled as the inertialess flow of a polymer melt \cite{Meulenbroek, black,origins_sharkskin}, and so the above results are relevant to this system. We emphasise the result that the growth rate of PDI diverges in the limit of small $\beta$ in units of $U_0/h$, meaning that polymer melts are especially susceptible to this instability. We discuss the connection between PDI and the `melt fracture' instabilities that are known to affect plastic extrusion in section \ref{discussion} and argue there that the effects of PDI are potentially significant.

\subsection{Flows with Inertia: $Re> 0$}\label{results_inertia}

We now introduce inertia and consider how this alters the instability. The analytic work demonstrated that $Re$ only enters the equations in the parameter group $\varepsilon Re / \beta$, and so this controls the strength of inertia. It will be useful to define 
$$ Re^* \coloneqq \frac{\beta}{\varepsilon}$$ 
to be the Reynolds number at which non-inertial behaviour transitions to inertial behaviour. The scalings and analysis already presented considered the limit of $\beta/W(1-\beta) \rightarrow 0$ when $ Re/Re^*$ is finite, however in this section we consider the effects of increasing $Re/Re^*$ with finite $\beta/W(1-\beta) \ll 1$. We use $Re/Re^* \sim 1000$ as the largest magnitude of this parameter. Bounding this parameter means it is finite, and hence provided the value of $\beta/W(1-\beta)$ is sufficiently small, the analytic predictions already presented remain valid. However, now that $Re/Re^*$ is no longer fixed, we incorporate it into our scalings to understand how $k$ and $c$ vary with it.

\begin{figure}[h!] 
\centering
    \begin{subfigure}[b]{0.49\textwidth}
    \centering
\includegraphics[width=\textwidth]{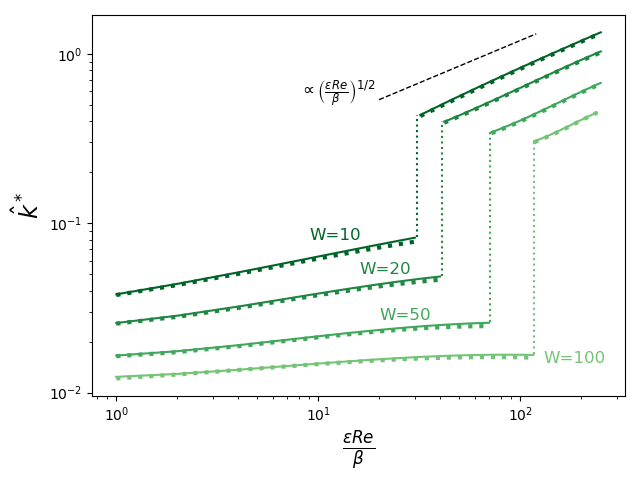}
    \caption{}\label{most_unstable_kx}
  \end{subfigure}
  \begin{subfigure}[b]{0.49\textwidth} 
    \centering
\includegraphics[width=\textwidth]{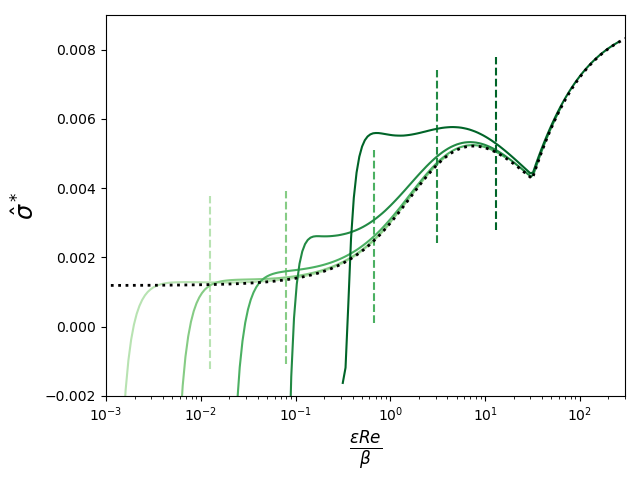}
    \caption{}\label{growth_rate_inertial}
  \end{subfigure}
\caption{a) Most unstable wavenumber $\hat k^*$ against $\frac{\varepsilon Re}{\beta}$, as $Re \rightarrow \infty$. We set $\beta=0.01$ and $\varepsilon=0.0001$, which sets $Re^* = 100$. $W=10,20,50,100$ (darkest to lightest), with numerics (solid lines) and predictions (dotted lines). b) The maximum growth rate against $\frac{\varepsilon Re}{\beta}$ as $\beta\rightarrow 0$ with numerics (solid lines) and predictions (dotted lines). Parameters are $W=10$, $\varepsilon=0.0001$, (from lightest to darkest), $Re= 10,40,160,640,2560$. The vertical dashed lines correspond to $\beta_{lim}(Re)$ which is explained in section \ref{validity}, and shows where the predictions are accurate to within $5\%$ of the true value. These plots show that when $\varepsilon Re / \beta$ is sufficiently large, PDI shifts towards inertial scalings of $\hat k \sim (\varepsilon Re / \beta)^{1/2}$, and $\hat \sigma \sim (\varepsilon Re / \beta)^{0}$.}
\end{figure}

\begin{figure}[h!] 
\centering
  \begin{subfigure}[b]{0.49\textwidth} 
    \centering
\includegraphics[width=\textwidth]{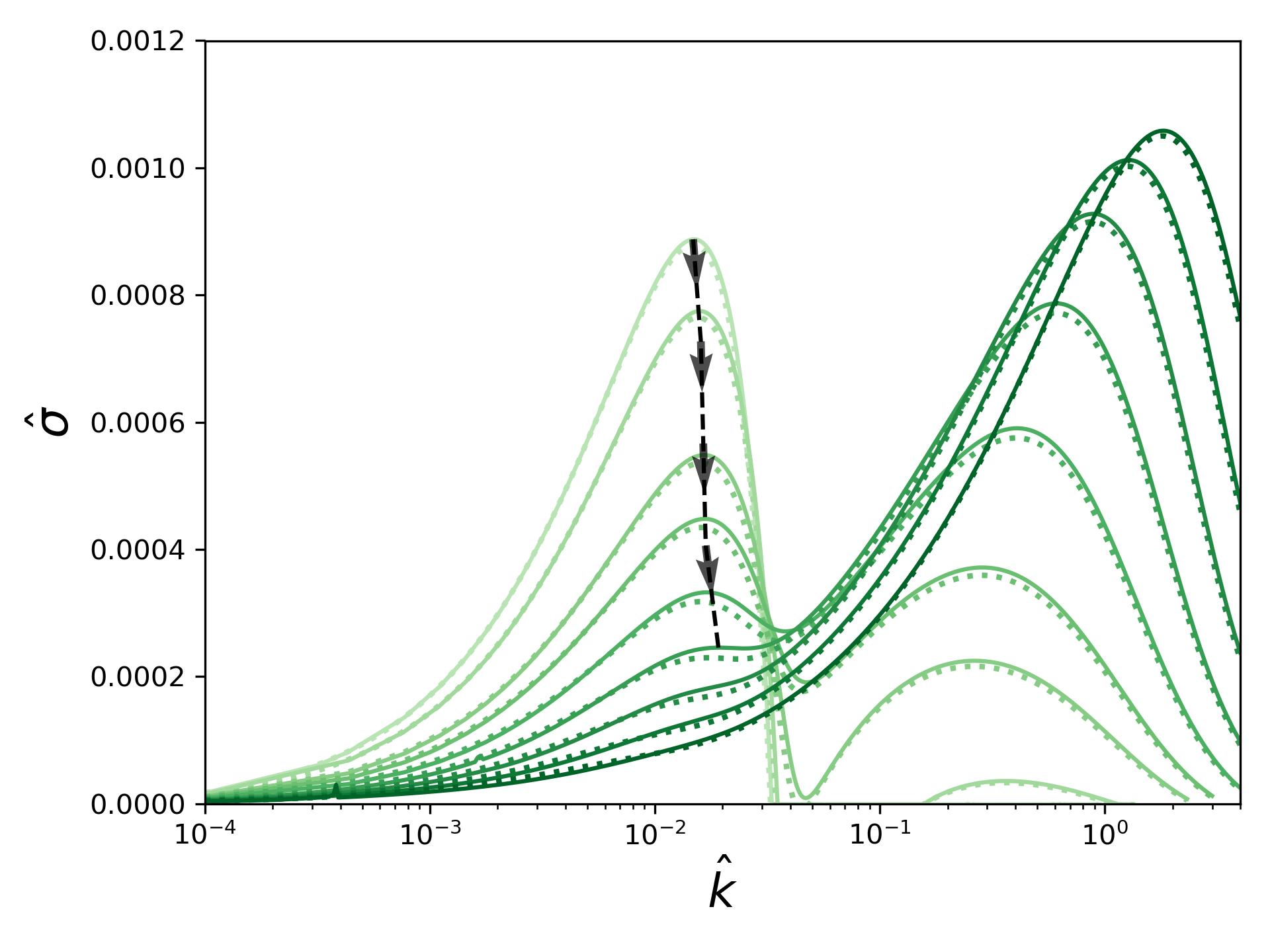}
      \caption{}
      \end{subfigure}
   \begin{subfigure}[b]{0.49\textwidth}
    \centering
\includegraphics[width=\textwidth]{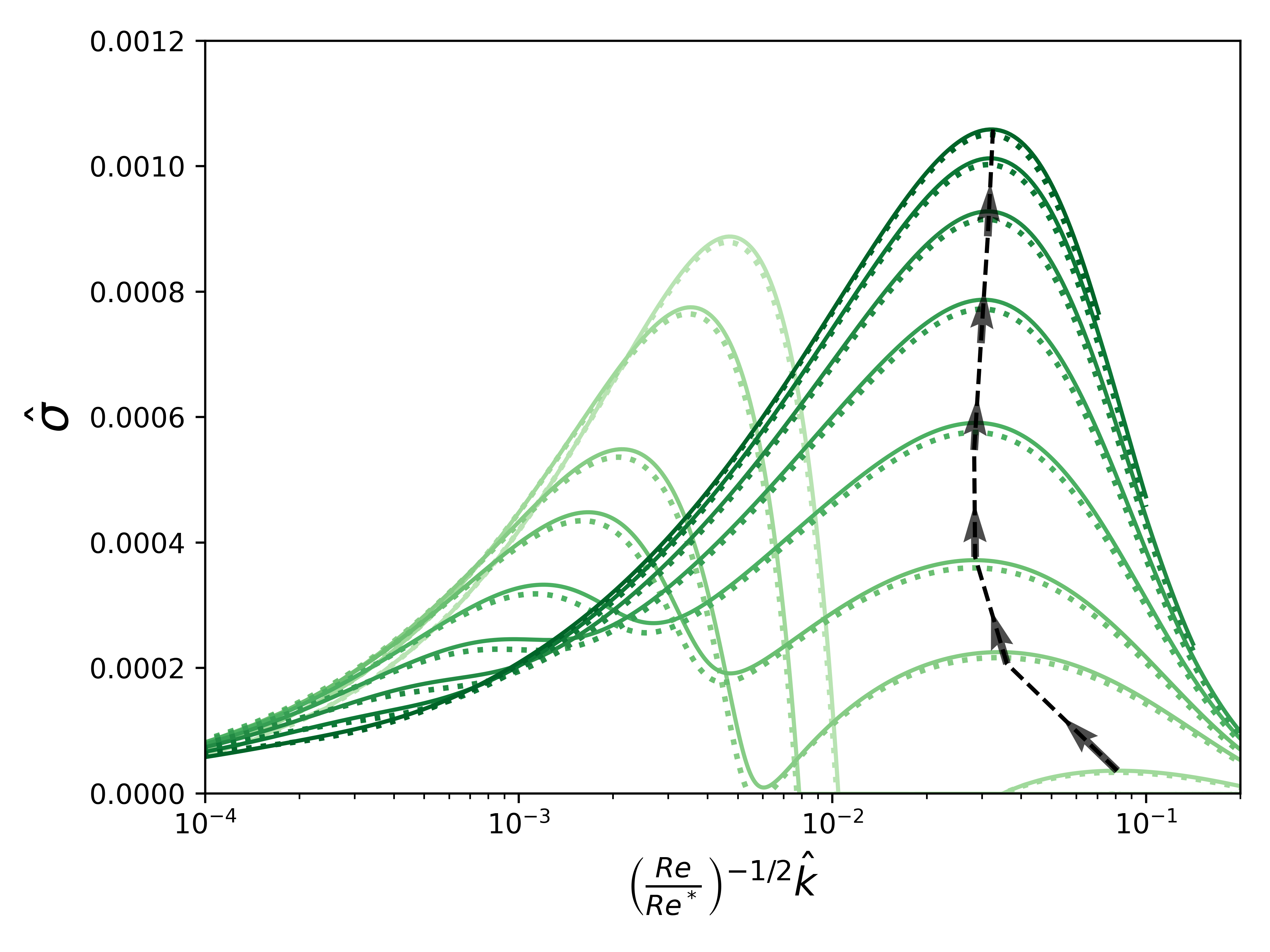}
\caption{}
      \end{subfigure}
\caption{Dispersion relations of growth rate against a) $\hat k$ and b) $(Re/Re^*)^{-1/2}\hat k$ obtained from our analytic predictions (dotted lines) and numerics (solid lines). Parameters are $W=100$, $\beta=0.01$, $\varepsilon=0.0001$ and (lightest to darkest) $Re = 1000, 2000, 6000, 10000, 20000, 40000, 80000, 160000, 320000$. These parameters set $Re^*=100$. The near-vertical black arrows connect the a) non-inertial first peaks and b) inertial second peaks with a cubic spline. Arrows point in the direction of increasing $Re$. This shows that the first peak occurs at non-inertial wavenumbers $\hat k \sim (Re/Re^*)^0$ and the growth in this region is suppressed as inertia increases. In contrast, the second peak is promoted by inertia, and exists at wavenumbers $\hat k \sim (Re/Re^*)^{1/2}$. The plots also show that $\hat \sigma \sim (Re/Re^*)^0$.
}\label{dispersion_relations_Re}
\end{figure}

In figure \ref{most_unstable_kx} we consider the relationship between the most unstable wavenumber $\hat k^*$ and $Re/Re^*$. The most obvious feature of this plot is that $\hat k^*$ is discontinuous, with it jumping to an inertial scaling of $\hat k^* \sim (Re/Re^*)^{1/2}$ at around $Re \sim 50 Re^*$. This transition point occurs at larger $Re/Re^*$ as $W$ is increased. This discontinuity is due to the eigenmode having different behaviours at $\hat k \sim (Re/Re^*)^0$ and $\hat k \sim (Re/Re^*)^{1/2}$. We examine this by considering the dispersion relation as we increase $Re/Re^*$ in figure \ref{dispersion_relations_Re}. As $Re/Re^*$ increases, the instability in the non-inertial region of $\hat k \sim (Re/Re^*)^0$ is suppressed, while it is promoted in the inertial region of $\hat k \sim (Re/Re^*)^{1/2}$. The discontinuity in $\hat k^*$ appears when the inertial region becomes more unstable than the non-inertial region. 

Figure \ref{growth_rate_inertial} shows the dependence of the maximum growth rate $\hat \sigma ^*$ on $Re/Re^*$, and it demonstrates that $\hat \sigma \sim (Re/Re^*)^0$. In the limit of $\beta \rightarrow 0$, for fixed $Re$, $W$ and $\varepsilon$, the maximum growth rate curves collapse when plotted against $Re/Re^* = \varepsilon Re/\beta$ showing that $\hat \sigma^* = \hat \sigma^*( Re/Re^*, W)$ in this limit, as suggested by the analytic predictions. The magnitude of this growth rate is not sensitive to the magnitude of $Re/Re^*$. A kink is seen in the maximum growth rate curves when the most unstable wavenumber transitions from non-inertial to inertial scalings.

\begin{figure}[h!] 
\centering
  \begin{subfigure}[b]{0.49\textwidth}
    \centering
\includegraphics[width=\textwidth]{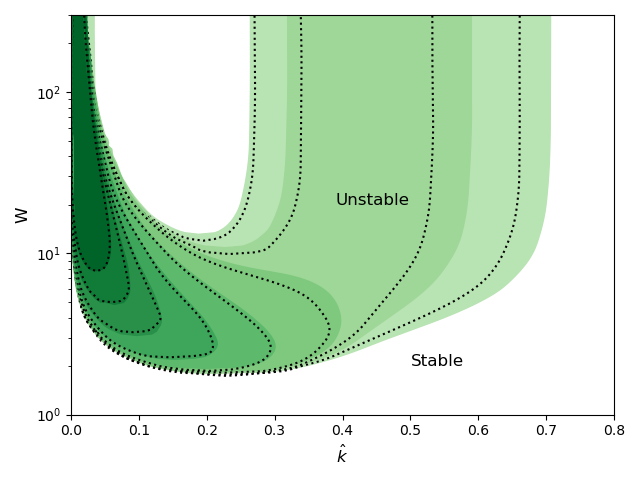}
    \caption{}\label{neutral_curve_Re1}
  \end{subfigure}
  \begin{subfigure}[b]{0.49\textwidth}
    \centering
\includegraphics[width=\textwidth]{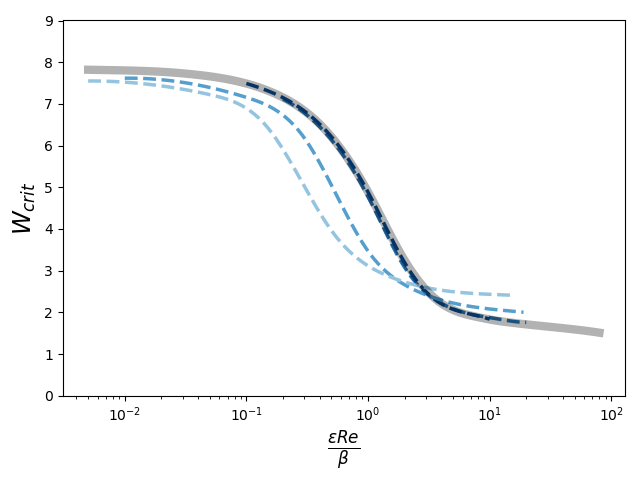}
    \caption{}\label{W_crit}
  \end{subfigure}
\caption{a) Neutral stability curves in the ($W$, $\hat k$) plane with changing $Re$ with $\varepsilon=0.0001$ and $\beta=0.01$. These set the transition from non-inertial to inertial behaviour to be at $Re^* = 100$. Darkest to lightest shade shows numerics for $Re=0,100,200,400,800,1200,1300,1400$. Dotted lines denote the neutral curve predicted by the analytic theory. Note when inertia is considered the prediction is less accurate - we expect that if a smaller $\beta$ were considered that the prediction would match the numerics more closely.  b) The relationship between $W_{crit}$ (the smallest $W$ at which the system goes unstable) and inertia. Each dashed curve tracks changing $Re$ with constant $\varepsilon = 0.0001$ and (darkest to lightest) constant $\beta = 0.001$, $0.01$, $0.1$, $0.2$. The transparent grey curve shows the analytic prediciton. } 
\end{figure}

Hence, for $Re/Re^* \gg 1$, the variables scale like
\begin{equation}
k \sim \frac{\sqrt{W (1-\beta)Re}}{\beta}, \quad c \sim \sqrt{\frac{W (1-\beta)}{Re}}
\end{equation}
Figure \ref{neutral_curve_Re1} shows how the neutral curve in the $(W,\hat k)$ plane changes when inertia is present. Increasing inertia destabilises the system, as more of the $(W, \hat k)$ plane becomes unstable. In particular, the smallest $W$ at which the system is unstable for some wavenumber (denoted by $W_{crit}$) decreases as $Re$ increases. As $Re$ is increased, the neutral curve starts to bend upwards at $Re \approx 13 Re^*$. This bend means that, for fixed $W$, there is a region of stable wavenumbers in between regions of unstable wavenumbers, as was seen for the $Re=2000$ curve in figure \ref{dispersion_relations_Re}. As $Re$ increases past $Re \approx 13 Re^*$, the unstable region of the neutral curve continues to expand, so that the stable region between the bends in the neutral curve gradually disappears (not shown).

In figure \ref{W_crit} we investigate how $W_{crit}$ changes with inertia, which is measured by the magnitude of $\varepsilon Re/\beta$. There are two distinct regimes determining how low $W_{crit}$ can be. When $Re \ll Re^*$ we get $W_{crit} \sim 8$, which is the same result obtained for $\beta \approx 0.5$ at $Re=0$ in \cite{beneitez}. However, when $Re \gg Re^*$, this critical value drops to $W_{crit} \sim 2$. This behaviour continues for viscosity ratios as large as $\beta=0.2$, as $W_{crit}$ still has a marked drop at $Re \sim Re^*$.

\begin{figure}[h!] 
    \centering
  \begin{subfigure}[b]{0.49\textwidth}
\includegraphics[width=\textwidth]{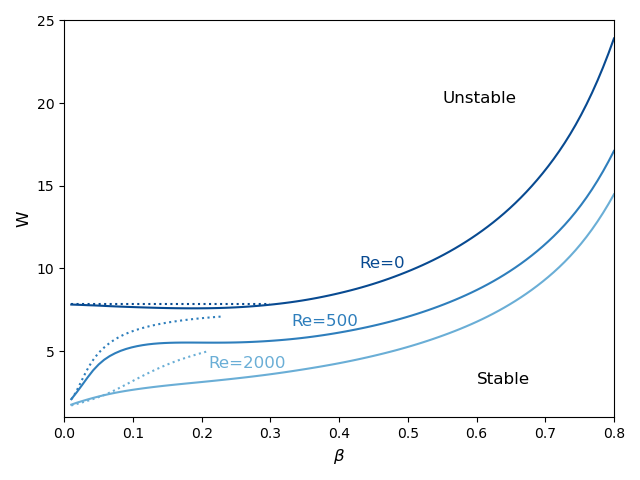}
  \end{subfigure}
  \begin{subfigure}[b]{0.49\textwidth}
\includegraphics[width=\textwidth]{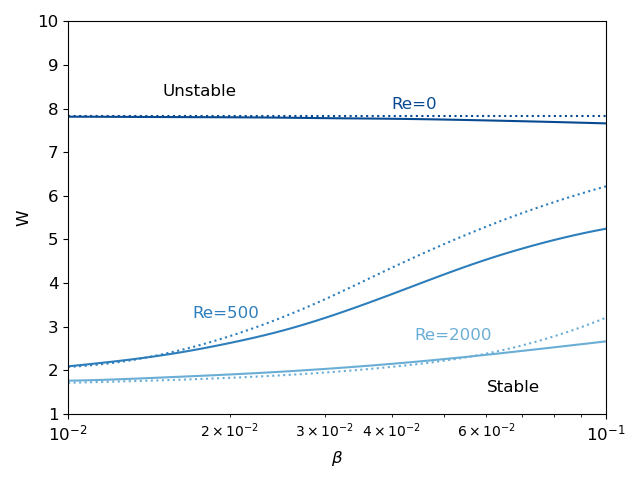}
  \end{subfigure}
\caption{The neutral curve in the ($W$, $\beta$) plane for plane Couette flow with diffusion-free boundary conditions at $\varepsilon=0.0001$ and $Re= 0,500,2000$. The right figure zooms in on the small $\beta$ region of the left figure, and uses a log scale. Solid lines correspond to numeric solutions, dotted lines correspond to the predictions based on the analytic results. Note each curve has constant $Re, \varepsilon$ so as $\beta \rightarrow 0$ we do not have ${\varepsilon Re}/{\beta}$ finite, which was required for the analytic work to be valid. This explains why the dotted prediction curves do not match the solid numeric curves exactly in the small $\beta$ limit. These plots demonstrate that PDI is generic across $\beta$, and that inertia destabilises the system.} \label{W-beta_neutral_curve}
\end{figure}

Lastly, we show neutral curves in the ($W$, $\beta$) plane in figure \ref{W-beta_neutral_curve} where we consider the stability across all wavenumbers. This shows that inertia generically decreases $W_{crit}$ across the whole range of $\beta$. In the small $\beta$ limit, this plot demonstrates again that $W_{crit} \sim 8$ when there is no inertia, but that including inertia lowers the critical Weissenberg number to $W_{crit} \sim 2$. The predicted neutral curves match the numerics as $\beta \rightarrow 0$. In the inertialess case of $Re=0$, the prediction for the inertialess case even appears to be accurate for viscosity ratios as high as $\beta=0.3$. These neutral curves demonstrate that plane Couette flow is susceptible to PDI over a large range of parameters.

\subsection{Prediction Validity}\label{validity}
\begin{figure}[h!] 
    \centering
  \begin{subfigure}[b]{0.49\textwidth}
\includegraphics[width=\textwidth]{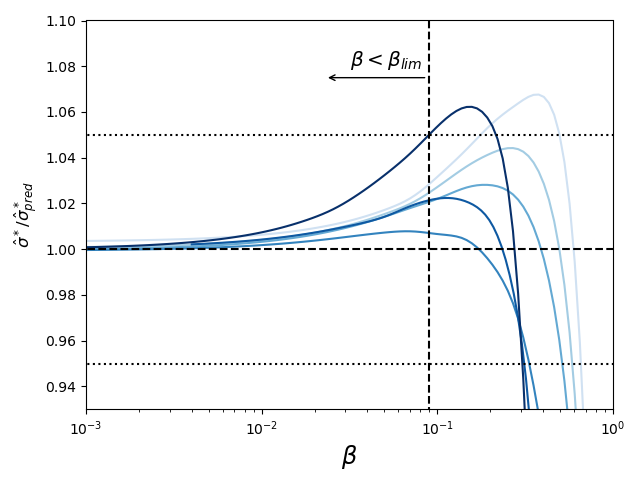}
    \caption{}\label{error_max_growth}
  \end{subfigure}
  \begin{subfigure}[b]{0.49\textwidth}
\includegraphics[width=\textwidth]{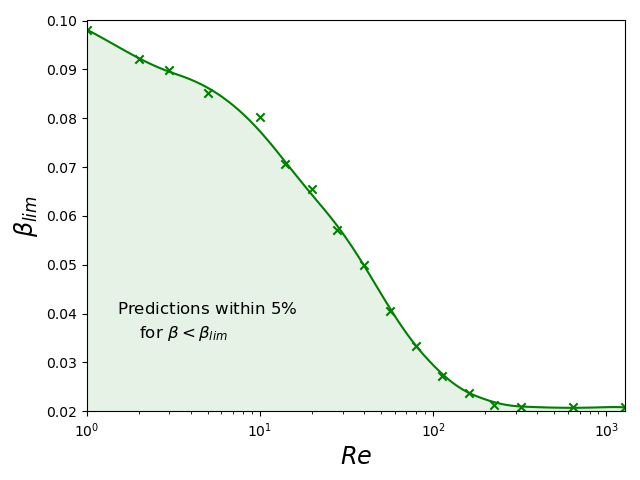}
    \caption{}\label{beta_lim}
  \end{subfigure}
\caption{a) The relative error in the predicted maximum growth rate $\sigma_{pred}^*$ as $\beta$ varies for $\varepsilon=0$, $Re=0$ and (darkest to lightest) $W=10, 12, 20,100,200,400$. For $\beta < \beta_{lim}(Re=0) = 0.090$ (vertical dashed line), all predictions are within $5\%$ of the true value (horizontal dashed lines). b) $\beta_{lim}$ as a function of $Re$. These plots verify that our predictions are asymptotically accurate, and demonstrate the size of $\beta$ at which the predictions become valid.} 
\end{figure}

We now quantify the accuracy of the predictions based on the maximum growth rate $\hat \sigma^*$. Figure \ref{error_max_growth} shows how the error in this prediction varies with $\beta$ for fixed $W, Re$ and $\varepsilon$. It demonstrates that the error decays as $\beta \rightarrow 0$, and it motivates defining $\beta_{lim}(Re)$ such that when $\beta < \beta_{lim}(Re)$, the predicted growth rate is within $5\%$ of the actual value. $\beta_{lim}$ is independent of $\varepsilon$, as the exact rescaled governing equations (\ref{rescaled:eq2}) and (\ref{rescaled:eq1}) are independent of $\varepsilon$ for plane Couette flow. To find $\beta_{lim}$ at a given $Re$, we consider the growth rate error curves for $W \in \{10,12,20,50,100,200,400\}$, and take $\beta_{lim}$ to be the largest $\beta$ such that all errors on all curves are smaller than $5\%$ for all $\beta < \beta_{lim}(Re)$. To illustrate this, figure \ref{error_max_growth} shows that $\beta_{lim}(Re=0) = 0.090$. Figure \ref{beta_lim} shows how $\beta_{lim}$ varies with $Re$, showing that our predictions require smaller $\beta$ to be valid as $Re$ increases. The plot suggests that any $\beta < 0.02$ is sufficient for predictions to be accurate to within $5\%$ for $Re<1000$.

\section{Robustness to Boundary Conditions}\label{robustness}

\begin{figure}[h] 
    \centering
\includegraphics[width=0.6\textwidth]{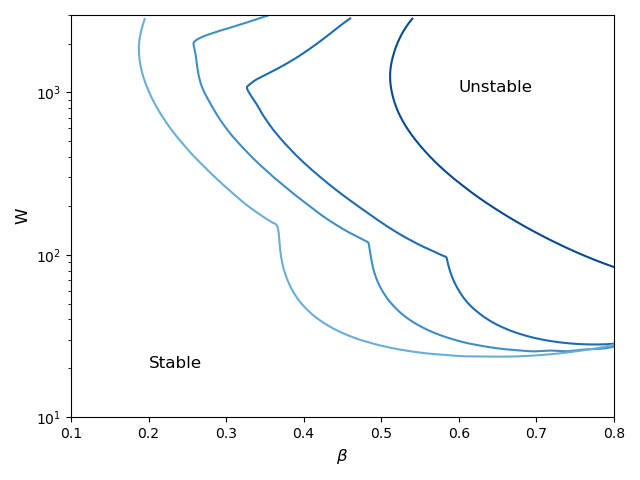}
\caption{Neutral curves in the ($W,\beta$) plane for Couette flow with Neumann boundary conditions. The other parameters are $\varepsilon=0.0001$ and, from darkest to lightest, $Re=0,500,1000,2000$.} \label{W-beta_couette_no_flux}
\end{figure}

So far we have considered the results of plane Couette flow with vanishing diffusion at the boundary ($\nabla^2\T|_{y=\pm1}=0$). We consider here how changing the boundary condition to vanishing stress flux ($D\T|_{y=\pm1}=0$) changes PDI, and we find that the instability is sensitive to this change. The laminar base flow does not change with the boundary condition in the case of plane Couette flow whereas a new boundary layer is introduced in the base profile for channel flow (see figure \ref{channel-base-flow}). We plot in figure \ref{W-beta_couette_no_flux} the neutral stability curve in the ($W, \beta$) plane for Couette flow with this Neumann boundary condition, and we see that PDI is suppressed in the small $\beta$ limit. There is now a minimum $\beta$ at which PDI exists, with PDI penetrating to lower $\beta$ when $Re$ is increased, showing more evidence that inertia destabilises the system. This means that in the concentrated limit of a polymeric fluid PDI does not exist for Neumann conditions on the polymer stress. There is also a maximum $W$ at which the instability exists. PDI is therefore sensitive to a change in boundary conditions. However, the diffusion-free boundary conditions seem  more likely to be physically relevant than the Neumann condition presented here but this, of course, is not certain.

\section{Discussion}\label{discussion}

\subsection{Connection to Melt Fracture}

Our work indicates that the polymer diffusive instability (PDI) is relevant for polymer melts where $\beta$ is very small, due to the divergence of the growth rate of PDI as $\beta \rightarrow 0$. PDI could therefore play a role in the melt fracture of extruded plastics. 
%
The `sharkskin' instability is a short wavelength distortion seen on the surface of extruded plastics that is associated with melt fracture \cite{Denn_wall_slip, sharkskin_experiment} and we now explore whether this could in fact be PDI.
%
%
We begin by comparing the observed wavelengths and those of the PDI. Figure 13a in \cite{sharkskin_experiment} shows the sharkskin wavelength found experimentally in a channel with half-width $h$ is in the range, $ \lambda_{sharkskin} \in (h/10, h) $. Using the non-dimensionalisation and notation introduced earlier, this corresponds to a wavenumber of $ k \in (2\pi,20\pi)$.  We will consider what value the polymer stress diffusion constant $\varepsilon$ should take for inertialess PDI to be unstable for such $k$. Taking $W \sim 10$, $\beta \sim 0.1$, then, from figure \ref{neutral_curves_creeping}, $\hat k \sim 0.04$ at the onset of instability. Since
$$ \varepsilon = {\frac{W(1-\beta)}{ \beta}} \frac{\hat k^2}{k^2}$$
then $\varepsilon \in (4\times 10^{-5}, 4\times 10^{-3})$, which is a reasonable value for the polymer diffusion to take \cite{beneitez}. Hence the wavelength at which PDI is unstable in a plastic extrudate is comparable to the wavelength of the observed sharkskin instability.

For PDI to be able to affect the extruding process, the growth rate must be large enough that any perturbation at the start of the process is sufficiently amplified by the time the plastic has exited the die. 
PDI is active at a lengthscale $\sqrt{\varepsilon \beta / W(1-\beta)}$ away from the boundary, and so the deforming material travels down the tube with velocity $U \sim \sqrt{\varepsilon \beta / W(1-\beta)}$. Hence the instability grows over timescale $T \sim L\sqrt{ W(1-\beta)/ \varepsilon \beta }$ where $L$ is the length of the die in units of half-channel height. The growth rate of PDI (from figure \ref{growth_rate_creeping}) is $\sigma \sim 0.001 W(1-\beta)/\beta$, and so the amplification of a perturbation over time $T$ is 
$$\exp(\sigma T) \sim \exp(0.001\varepsilon^{-1/2}W^{3/2}(1-\beta)^{3/2}\beta^{-3/2}L).$$
Taking the parameter values, $\varepsilon \sim 10^{-3}$, $W \sim 5$, $\beta \sim 0.1$ and $L \sim 10$, yields a huge amplification of $3 \times 10^{41}$.

\begin{figure}[h!] 
\centering
\includegraphics[width=0.6\textwidth]{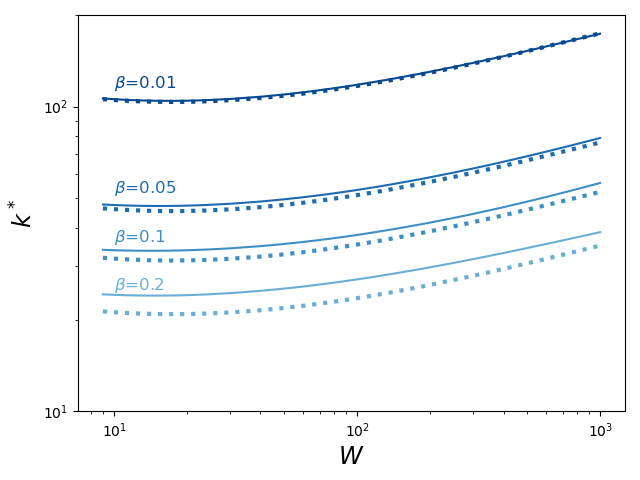}
\caption{The most unstable wavelength, $k^*$ against $W$ for $Re=0$, $\varepsilon=0.0001$ and $\beta=0.01,0.05,0.1,0.2$. Numerics are shown by the solid lines, predictions based on the analytic theory are shown by dotted lines.} \label{most_unstable_kx_creeping}
\end{figure}

However, while clearly strong enough, PDI does not reproduce the relationship between wavelength and shear rate that is seen for the sharkskin instability. Specifically, the wavenumber of sharkskin is seen to decrease with shear (see figure 13a in \cite{sharkskin_experiment} again), while the most unstable wavenumber associated with PDI increases with shear (see figure \ref{most_unstable_kx_creeping}).

%
%

\subsection{Physical or artificial?}

We now discuss the possibility that PDI may be an artificial instability. This is an issue because the lengthscale of the boundary layer obtained for PDI approaches that of the polymer lengthscale, suggesting that the continuum assumption is not obviously valid \cite{beneitez}. We consider the largest PDI lengthscale, which comes from the inertialess case (the most unstable wavenumber grows with inertia in figure \ref{most_unstable_kx}) and so this case is most likely to provide the required separation of scales for the continuum assumption to be reasonable. 

Figure \ref{most_unstable_kx} suggests that the most unstable wavenumber for inertialess PDI occurs at $\hat k^* \approx 0.04$, and hence dimensionally we have a most unstable wavelength of $ 150 \sqrt{\varepsilon\beta/W(1-\beta)} h$. As in \cite{beneitez}, the Stokes-Einstein relation is used to consider the ratio of the PDI boundary layer lengthscale to the polymer gyration radius $R$ for parameters relevant to low density polyethylene (LDPE), which is known to show signs of the sharkskin instability \cite{Denn}. 
At temperatures of $T \sim 500K$, LDPE has $\mu_s \sim 100 Pa\cdot s$, $R \sim 10 \,nm$ \cite{LDPE_gyration_radius} and its relaxation spectra consists of a range of relaxation times as large as $\lambda \sim 100\,s$ and as small as $\lambda \sim 10^{-4}s$ \cite{LDPE_relaxation_time}. For $W\sim10$ and $\beta \sim 0.1$ the PDI lengthscale is therefore
$$150 \sqrt{\frac{\varepsilon\beta}{W(1-\beta)}} h \sim  \sqrt{\frac{\beta}{(1-\beta)}} \frac{150R}{W} \sqrt{\frac{k_B T \lambda}{6\pi \mu_s R^3}} \in  (0.1R, 100R)$$
where $k_B$ is the Boltzmann constant. This shows that the lengthscale of PDI is sensitive to the choice of $\lambda$. Large $\lambda$ makes the lengthscale of PDI much larger than $R$, meaning the continuum assumption is valid as there is the required separation of scales. However, small $\lambda$ makes PDI act on a lengthscale smaller than $R$, invalidating the continuum assumption. Due to this, it is unclear how physical PDI is for polymer melt systems.

\subsection{Conclusion}

The polymer diffusive instability is analytically tractable in the limit of $\beta/W(1-\beta) \rightarrow 0$ for finite ${\varepsilon Re}/{\beta}$. The eigenvalues of PDI can be predicted when boundary conditions of vanishing diffusion are used. We find $\beta_{lim}(Re)$ such that when $\beta < \beta_{lim}(Re)$, a prediction of the maximum growth rate is within $5\%$ of its true value, and we see that $\beta < 0.02$ yields accurate predictions for $Re<1000$. 

In the limit of $\beta/W(1-\beta) \rightarrow 0$, the smallest $W$ at which an instability is seen ($W_{crit}$) depends on the strength of inertia, measured by $\varepsilon Re/\beta$. Inertia is shown to destabilise the system, with it allowing the system to become unstable at lower $W$ than in the inertialess case. When $\varepsilon Re/\beta < 0.1 $, we obtain $W_{crit} \sim 8$, while when $\varepsilon Re/\beta > 5 $ we obtain $W_{crit} \sim 2$. Similar behaviour is even seen outside of the concentrated limit, when $\beta \sim 0.2$. PDI is sensitive to a change in boundary conditions however, as changing the diffusion-free boundary conditions to vanishing stress flux boundary conditions changes where in parameter-space the instability is operative. In particular, this change suppresses PDI in the small $\beta$ limit. 

PDI seems particularly relevant to the case of polymer melts since the growth rate of PDI is so large. This indicates it could be relevant to real extrusion devices where the most unstable wavenumber is similar to the wavenumber of the observed `sharkskin' instability for typical parameters.
%
However, in some parameter ranges, the wavelength over which PDI operates approaches the lengthscale of the polymers themselves which invalidates the continuum viscoelastic model used (here Oldroyd-B but also FENE-P as discussed in \cite{beneitez}).
We are therefore left to contemplate whether a `better' continuum model would suppress PDI in these troublesome parts of parameter space where PDI may not be physical. Hopefully experiments can help resolve this as well as studies using more sophisticated continuum models. We hope to report on the latter (at least) in the near future.

\FloatBarrier 

\newpage

\appendix

\section{Channel flow Results}\label{analogous_results}

In the main text we considered the results for plane Couette flow with no diffusion at the boundary ($\nabla^2 \vb T|_{y=\pm1}=0$). Then in §\ref{robustness} we considered the neutral curves for plane Couette flow when vanishing stress flux is used instead ($D \vb T|_{y=\pm1}=0$), and we saw that this suppressed PDI in the small $\beta$ limit. Here we present the analogous results for channel flow.

\subsection*{Diffusion-Free Boundary Conditions ($\nabla^2 \vb T|_{y=\pm1}=0$)}\label{diffusion-free-channel}

We plot in figure \ref{Wi-beta_poiseuille_no_diff} the neutral curve in the ($W,\beta$) plane. It is almost identical to the neutral curve in the plane Couette case, with them looking the same by eye. This is because PDI in this system is sufficiently localised at the boundary, where the plane Couette and channel flow base flows appear identical.

\subsection*{Neumann Boundary Conditions ($D \vb T|_{y=\pm1}=0$)}\label{neumann-channel}

\begin{figure}[h!] 
\centering
  \begin{subfigure}[b]{0.45\textwidth} 
\includegraphics[width=\textwidth]{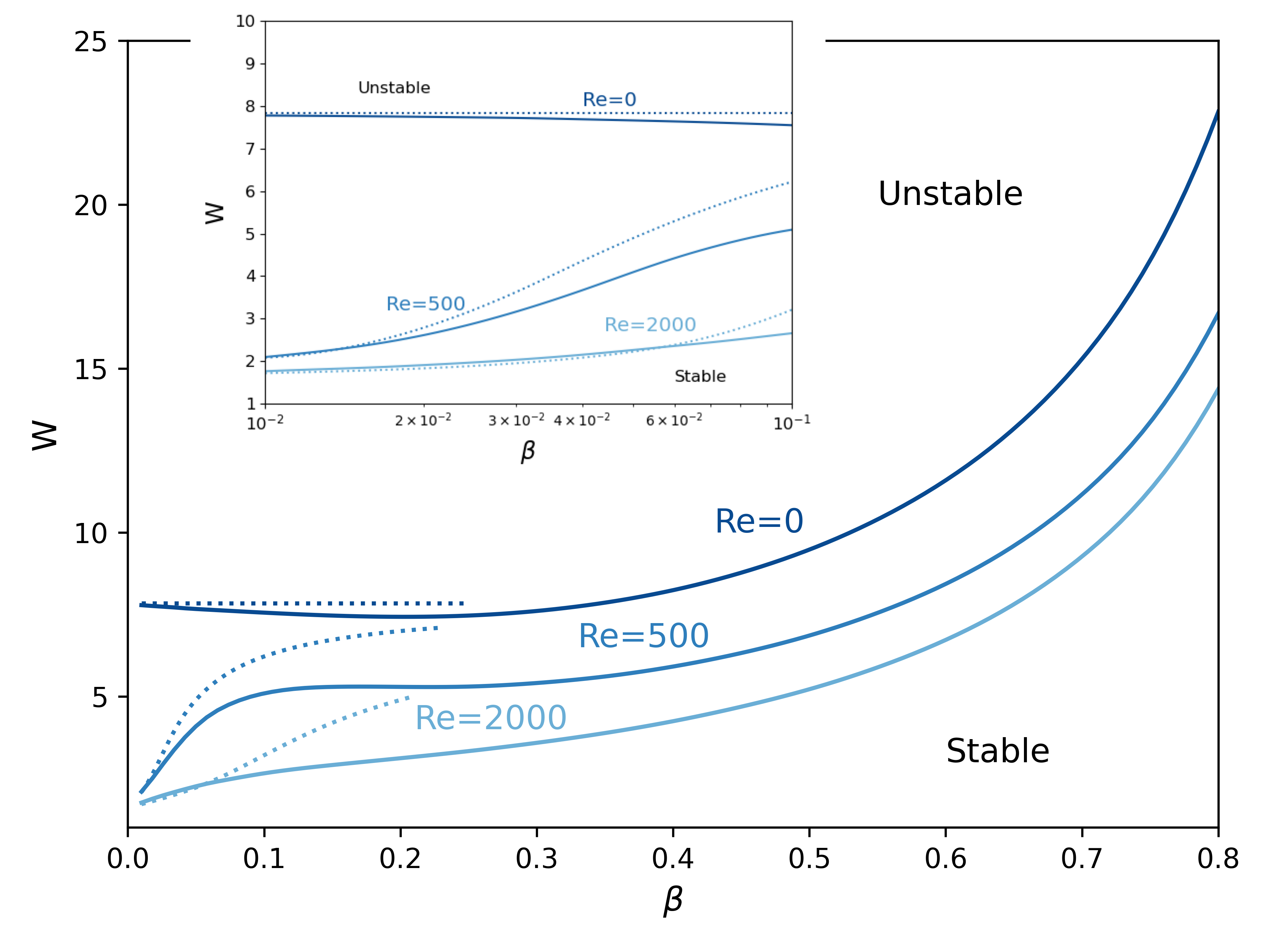}
\caption{} \label{Wi-beta_poiseuille_no_diff}
  \end{subfigure}
  \begin{subfigure}[b]{0.45\textwidth} 
\includegraphics[width=\textwidth]{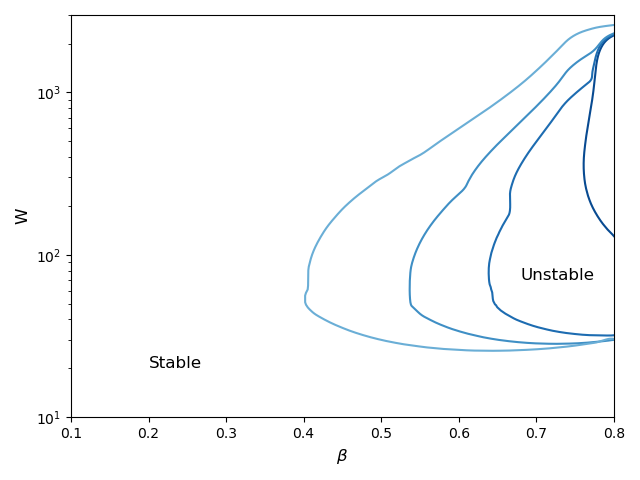}
\caption{} \label{Wi-beta_poiseuille_neumann}
  \end{subfigure}
\caption{a) The neutral curve in the ($W$, $\beta$) plane for channel flow with no diffusion at the boundary at $\varepsilon=0.0001$ and $Re= 0,500,2000$. Solid lines correspond to numeric solutions, dotted lines correspond to the predictions based on the analytic results. Inset shows neutral curve at small $\beta$ with log scale. Note that each curve has constant $Re, \varepsilon$ and hence in the small $\beta$ limit we do not have $\varepsilon Re/\beta$ finite, which was required for the analytic work to be valid. This explains why the dotted prediction curves do not match the solid numeric curves exactly in the small $\beta$ limit. b) Neutral curve in the ($W$, $\beta$) plane for channel flow with Neumann boundary conditions with $\varepsilon=0.0001$ and, from darkest to lightest shade, $Re=0,500,1000,2000$.}
\end{figure}

Like for plane Couette flow, changing to Neumann boundary conditions changes the stability of the system drastically in the small $\beta$ limit. We note that the change in the base state (see figure \ref{channel-base-flow} in the main text) is not solely responsible for this change, as in plane Couette flow the laminar base flow does not change with the boundary conditions, but a change of stability is still seen. We plot in figure \ref{Wi-beta_poiseuille_neumann} the neutral curve in the ($W, \beta$) plane. As was the case for plane Couette flow, the stability of the system is sensitive to a change in boundary conditions. In particular, PDI is suppressed at small $\beta$ with this change of boundary conditions. This means that the stability of a polymer melt in channel flow changes depending on the boundary conditions used.

\bibliographystyle{elsarticle-num} 
\bibliography{bibliography}
\end{document}